# Pavideoge: A Metadata Markup Video Structure in Video Search Engine

Pu YANG, Jun GUO, Guang CHEN
School of Information and Telecommunication Engineering
Beijing University of Posts and Telecommunications, Beijing, China
100876

puyang.bupt@gmail.com, {guojun, chenguang}@bupt.edu.cn

## ABSTRACT
In this paper, we study the problems of video processing in video search engine. Video has now become a very important kind of data in Internet; while searching for video is still a challenging task due to the inner properties of video: requiring enormous storage space, being independent, expressing information hiddenly. To handle the properties of video more effectively, in this paper, we propose a new video processing method in video search engine. In detail, the core of the new video processing method is creating pavideoge—a new data type, which contains the video advantages and webpage advantages. The pavideoge has four attributes: real link, videorank, text information and playnum. Each of them combines video's properties with webpage's. Video search engine based on the pavideoge can retrieve video more effectively. The experiment results show the encouraging performance of our approach. Based on the pavideoge, our video search engine can retrieve more precise videos in comparison with previous related work.

## Categories and Subject Descriptors
H.3.3 [**Information Storage and Retrieval**]: Information Search and Retrieval – *Information filtering, Search process.*

## 1. INTRODUCTION
Video is the technology of electronically capturing, recording, processing, storing, transmitting, and reconstructing a sequence of still images representing scenes in motion [1]. Video attracts more and more people to enjoy because video has unparalleled advantages: strong visual impact and beautiful voice. However, it is difficult for people to watch various videos easily due to the tremendous volume of video data. In order to solve this problem, with the vigorous development of the Internet, a number of online video sites are established to share all kinds of videos. Thus, people feel convenient and comfortable to enjoy real-time videos. An investigation made by us shows that almost all users who access to the online video sites want to find the precise video they want to watch. Nevertheless, with increasing the number of online video sites and video data on them, it is impossible to find the desired precise videos in the wide-area Internet. Thus, video search engine has become popular. A video search engine is a web-based search engine which crawls the web for video content [2]. Most popular commercial search engines such as Google [3], Baidu [4], and Sogou [5] have launched video search products.

However, the commercial video search engines [3][4][5] which basically adopt the traditional video processing method are usually ineffective in retrieving videos. This is because video has some inherent properties. First, the video requires enormous storage space; therefore it is hard for crawler to download all video data in real time. The investigation made by us also shows that about 85% users who access to the online video sites want to download some favorite videos. Because the videos are stored in video database in online video sites, it is impossible for users to download directly from the user interface. The traditional video processing method takes no action about this problem. Second, all of the video data are independent of each other. They cannot be dealt with by search engine. The traditional video processing method in video search engine uses tags to solve this problem. All of the commercial video search engines use this method to connect videos with other different videos surrounding them. Unfortunately, because these tags provide search engines with little information, the results from the search engines based on tags are not precise. Third, videos express information hiddenly. Webpages convey information via text, while all of the video data need special decoder to express information. The decoders are different due to the various types of video format and it cost a long time to decode videos. Using the video decoders in video search engine, the video search engines get results slowly. The traditional video processing method in video search engine uses titles to solve this problem. All of the commercial video search engines use this method to distinguish videos from the other videos. Unfortunately, because these titles contain little information, the search engines cannot get precise results. Because the traditional video processing method does not handle these immanent properties of video effectively, this motivates us to explore a better video processing method in video search engine.

In this paper, we propose a novel video processing method for retrieving video effectively. First, we collect data and create pavideoge—a new data type. "Pavideoge" indicates the mixture of page and video. Just like its name, pavideoge contains the video advantages and webpage advantages. The pavideoge has four attributes: real link, videorank, text information and playnum. Each of them combines video's properties with webpage's. Moreover, we use these four attributes to deal with the properties of video and enhance the performance of the video search engine. Real link, videorank and text information are used for settling the three intrinsic properties of video. Playnum is used for improving results' precision. For video requiring enormous storage space, our novel method provides pavideoge's real link for users to download video whenever they want to, while the traditional means do nothing. For video being independent, we propose a novel concept—videorank to connect each video with the others. Videos with videorank build virtual networks as webpages with hyperlinks and pagerank make up the real networks. Videorank can be used for improving the search results' precision because of its wealth information. For video expressing information hiddenly, we give video text information which comes from playpage. The playpage is a special webpage which is used for playing video online. The playpage consists of all kinds of text information about the video it plays, such as title, tags, relevant videos' titles, users' comments and so on. All of these can help search engine improve the precision.

The rest of this paper is organized as follows. First, we briefly review some related works in Section 2, and define the problem settings in Section 3. The system overview of the proposed approach is introduced in Section 4, and the algorithm details are described in Section 5. Experiment evaluations are reported in Section 6. And in the last section, we draw conclusions and point out some future research directions.

## 2. RELATED WORK

To make a tradeoff between performance and cost, all commercial video search engines based on traditional video processing method adopt simple match strategy.

There are two types of engine based on the traditional video processing method. One type of engines just uses title to match the search keywords. The typical video search engine of this type is Google's video search engine [3]. The Google's video search results indicate that Google's search engine simply crawl the video's titles from the online video sites and use these titles instead for video's content to match the search keywords (see Figure 1). Another type of engines utilizes title and tag to match. Baidu's video search engine [4] and Sogou's video search engine [5] are two typical instances. From the results (see Figure 1 and Figure 2), we can see that the three video search engines based on the traditional video processing method retrieve a few relevant videos. Video contains lots of information. Title and tag have so little information that they cannot fully express video content.

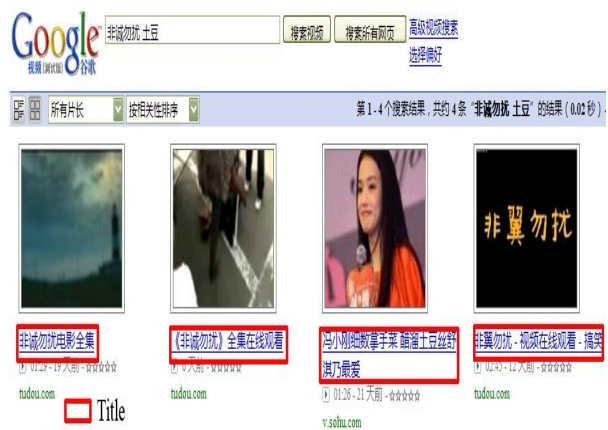

**Figure 1. One of the search results of Google's video search engine. Characters in red boxes are titles.**

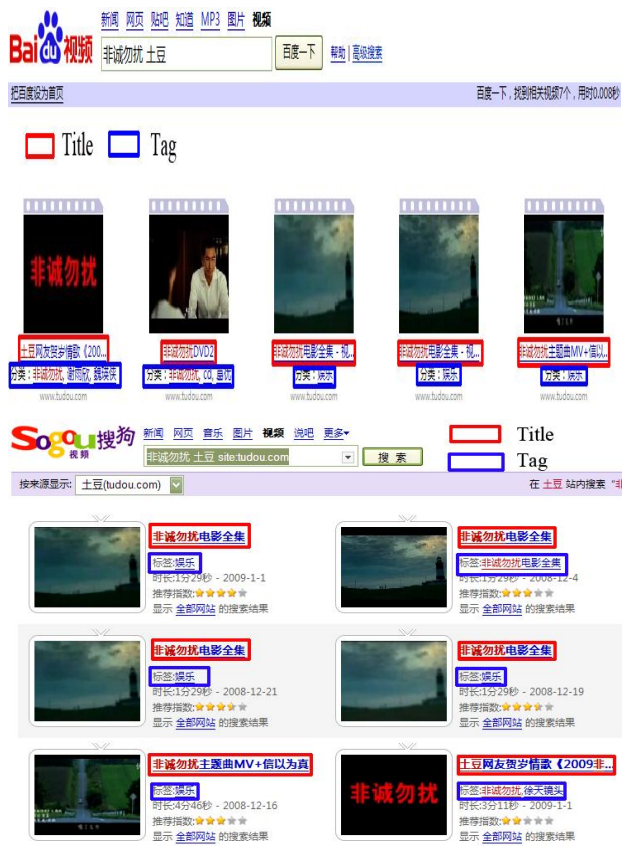

**Figure 2. One of the search results of Baidu's video search engine and one of the search results of Sogou's video search engine. Characters in red boxes are titles and Characters in bule boxes are tags.**

## 3. PROBLEM SETTING

To make a clear presentation and to facilitate the following discussions, we first explain three inner properties of video used in this paper.

## 3.1 Requiring Enormous Storage Space

Video contain much more information than webpage. It obviously requires more storage space than webpage. In our previous research, we randomly crawled about 80,000 videos and videos' webpages from tudou [6], one of the biggest online video sites in China. After we doing statistics, results showed that different videos had different space, while the different webpages had almost same size and the video's storage space was 1,000 times as large as webpage's on average. Thus, it definitely costs a lot of time to download videos for crawler of video search engine and it certainly decreases the real-time performance of video search engine. Therefore, we need a method instead of downloading videos.

## 3.2 Being Independent

Videos, not like webpages, have no hyperlinks. All of the video data are independent of each other in Internet without the help of other indicators, such as titles, tags and so on.

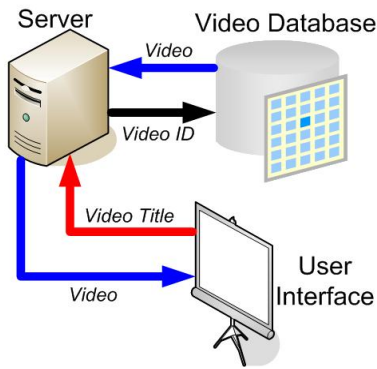

**Figure 3. An illustration of getting video in online video site**

Even in the online video sites, videos are also independent. Figure 3 illustrates how to get video in online video site. First, user clicks the title of the video through user interface such as web browser. At the same time the user interface sends video's title to the server of online video site. Then the server uses this title to find the video ID. Video ID is a series of characters which are used by online video site to distinguish one video from the other videos. Moreover, the server gets the video in video database via video ID. Finally, the server sends the video to user interface. In Figure 3, we can see that in online video sites, only title and its video are linked by video ID, while videos have no relationship between each other. Because the videos themselves have no connection between each other, it is hard for video search engine to retrieve more precise videos for users. The commercial video search engines [3][4][5] based on traditional method use titles' relationship sometimes with tags instead of videos' relationship. Although it may work, the relationship of titles and tags is far from enough.

## 3.3 Expressing Information Hiddenly

Not like the webpages，videos cannot be decoded by web browser without other special decoders. Notice that videos in the online video sites are handled by a special decoder called flashplayer in general. Compared with webpage, video needs much more time to be decoded. Additionally, video needs special tools to abstract information, which will certainly cost lots of time. Thus, video expresses information hiddenly. The commercial video search engines [3][4][5] based on traditional method only use title's and tag's text information instead of videos' information. Unfortunately, because these titles and tags contain much less information than videos' content, the search engines cannot retrieve more precise results. Therefore, we need a method which can get more useful information than titles and tags without decoding videos.

## 4. SYSTEM OVERVIEW

The flowchart of our method is illustrated in Figure 4, which mainly consist of two parts: (I) building pavideoge for every video and (II) creating search result.

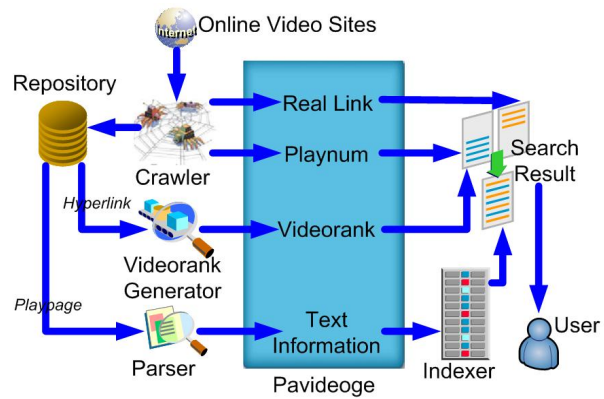

**Figure 4. The flowchart to the proposed approach**

In our novel video processing method, the goal of the first part is to gather useful data from online video sites and build pavideoge for every video by using these data. First, we crawl playpages and abstract hyperlinks in them. In addition, we save the playpages and hyperlinks orderly in repository for further use. Moreover, the hyperlinks are used to obtain pavideoge's real link by analyzing them in online video sites' environment. Furthermore, we also get pavideoge's playnum via analyzing playpage's hyperlinks. The above operations are driven by crawler in Figure 4. Second, Videorank generator (see Figure 4) utilizes the hyperlinks to calculate video's videorank. The videorank is discussed in detail in the next section. Third, we utilize the playpage previously saved in repository to create pavideoge's text information by parser, which is shown in Figure 4. Finally, we build pavideoge database by acting above three steps. More details of this part are discussed in section 5.

The second part is to create search result based on pavideoge database. First, we use pavideoge's text information which contains rich information about video to build index based on language model. Indexer is shown in Figure 4. Second, we use pavideoge's real link to indicate the video's downloading URL with video's online playing URL, as shown in the upper right of the Figure 4. Third, we utilize playnum and videorank to modify the search result in order to enhance the quality of the search result. This part of the method is discussed in detail in the following section.

## 5. METHOD BASED ON PAVIDEOGE

In this section, we present the details of the algorithm for processing video in online video sites for video search engine. First, the pavideoge is built for every video. Second, the search result is modified.

## 5.1 Building Pavideoge

To build pavideoge, all we need to do is to generate pavideoge's four attributes: real link, playnum, videorank and text information.

### 5.1.1 Real Link

After investigating the organization structures of a substantial number of various online video sites, we found that online video sites have the following two characteristics (see Figure 3).

- Server and video database are laid out separately in online video site.

- Server use video ID to get video in video database.

Therefore, we could get video's real link just by simulating the actions the server takes to obtain videos in a video database of online video site.

**Figure 5. The slice of source code of the playpage. Characters in blue box are video ID.**

To make this clear, we use an example to illustrate. We randomly select one video's URL in tudou [6]: http://www.tudou.com/programs/view/mCZ03uY6zYM/. First, we get playpage through the URL. The slice of source code of the playpage is shown in Figure 5. And then we find video ID in the source code (see blue box in Figure 5). Furthermore, we use the webpage called by the flash player of the online video site. Different online video sites have different formated webpages called by the flash player. In tudou [6], the format is: http://www.tudou.com/player/v.php. Finally, we simulate server's action, that is, using video ID to get video. By fetch http://www.tudou.com/player/v.php?id=11272862, we can get real link of the video (see Figure 6).

**Figure 6. The real link of video. Characters in blue box are real link.**

### 5.1.2 Playnum

Playnum is the number of times a video is played. It is one of the critical video's properties. It indicates whether the video is important and popular or not. In online video sites, playnum is dynamically generated by application in server and it cannot be obtained from the playpage downloaded by the crawler. Thus, we also need to simulate the server to obtain the video's playnum. We use the video's URL mentioned above to illustrate how to obtain playnum. First, we also need to get the video ID and the steps are the same as above. Second, when we get the video ID—11272862, we also simulate the server by using server's viewpage to get playnum. In this example, we fetch the http://www.tudou.com/programs/view_ajax.php?itemID=11272862 from online video site to get playnum. The fetched webpage is shown in Figure 7.

**Figure 7. The playnum of video. Characters in blue box are playnum.**

### 5.1.3 Videorank

Videorank is used to reveal the immanent regulation of the online video site, and it is generated when the video site is built. Different video sites have the different regulations due to adopting different building algorithms. Videorank is similar to pagerank [7][8][9][10][11][12], which is defined as follows:

Assume page A is pointed by pages T1...Tn (i.e., are citations). The parameter d is a damping factor between 0 and 1. We usually set d to 0.85. Also C(A) is defined as the number of links going out of page A. The pagerank of A is given as follows [7]:

$$PR(A) = (1-d) + d\left(\frac{PR(T1)}{C(T1)} + ... + \frac{PR(Tn)}{C(Tn)}\right) \qquad (1)$$

In pagerank, all of the hyperlinks pointing to the webpage should be calculated. However, it is not appropriate in online video sites. The playpage in online video sites contains lots of useless hyperlinks, such as advertisements, images, users' profile, webpages of other sites and other playpages which have no relationship with this playpage recommended by video sites. All of these are harmful to reveal the objective distribution of the videos' connection, which is generated by video site according to building algorithm.

In order to reveal the videos' distribution correctly, we modify the pagerank algorithm and introduce vidorank. Videorank is defined as follows:

Assume that video V's playpage is P, and P is pointed by playpages P1...Pm. As mentioned above, we set d to 0.85. Also

C(Pi) is defined as the number of playpage' links going out of playpage Pi. The videorank of V is given as follows:

$$VR(P) = (1-d) + d\left(\frac{VR(P1)}{C(P1)} + ... + \frac{VR(Pm)}{C(Pm)}\right) \quad (2)$$

$$VR(V) = VR(P) \quad (3)$$

In Figure 7, the areas enclosed by blue boxes indicate the playpages P1...Pm of the video V's playpage P. The list of P and Pi are shown in Figure 8. We can see all of URLs are playpages of tudou [6].

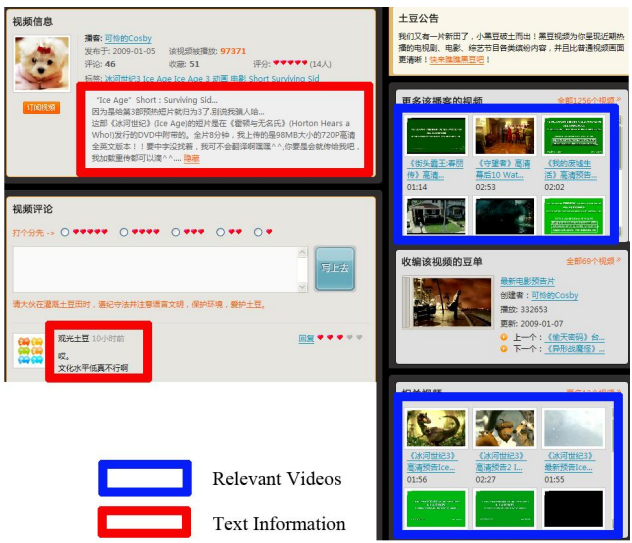

**Figure 7. The playpage of a video. The screen shot has been simplified for a clear view.**

### 5.1.4 Text Information
The playpage of the video contains lots of text information that describe the video's content, such as title, tags, relevant videos' titles, users' comments and so on. All of these can help search engine retrieve more precise results. In Figure 7, the area enclosed by the upper red box is the title, tags, and video description. The lower red box encloses the uers' comments after watching the video. The relevant videos' titles are inside the two blue boxes. All of these are marked by HyperText Markup Language (html). Thus, we could abstract them from the source code of playpage, delete the marks of html and save as text. Finally, we get the pavideoge's text information.

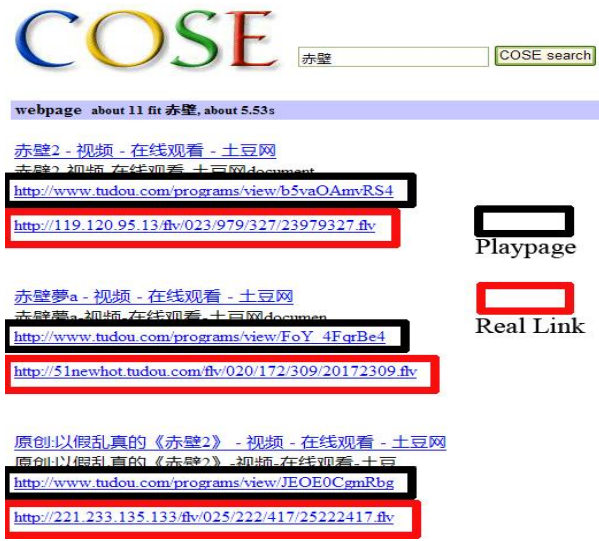

**Figure 8. The list of playpages of a video**

## 5.2 Modifying Search Result
After building index using text information based on language model, we use the other three pavideoge's attributes to modify search result.

### 5.2.1 Modified by Real Link
Real link is the hyperlink of the video in video database. User can directly download the video through real link. We add real link in search result with the hyperlink of playpage in order to meet the needs of the users and provide the solution to handle the video's enormous storage space. One of the search result modified by real link is shown in Figure 9.

**Figure 9. One of the search result modified by real link. The screen shot has been simplified for a clear view.**

### 5.2.2 Modified by Playnum and Videorank
Playnum is the unique attribute of the video. The playnum of the video in online video site indicates the popular and important of the video from the users' perspective. However, videos may be clicked by mistake and the users open the video but they do not watch it. When the above condition happens, the importance of the playnum is affected. Thus, we need an effect factor to control its proportion of affecting research result.

Videorank is used to reveal the immanent regulation of the online video site. It use the relevant videos generated by online video sites according to certain building algorithms to calculate the importance of the video. It is the factor of affecting search result from the online video sites' perspective. Nevertheless, the building algorithms adopted by video sites are not the most fitful for the video sites. Maybe the video were very important and popular, but the video were not recommended by video site due to the building algorithms. Thus, we need an effect factor to control its proportion of affecting research result either.

In order to make the modifying algorithm clear, we first introduce the definition of language model [13][14] used in our method. Language model is defined as follows [13]:

$$p(q|d) = \prod_i p(q_i|d)$$
$$= \prod_{i:c(q_i;d)>0} p_s(q_i|d) \bullet \prod_{i:c(q_i;d)=0} p_u(q_i|d)$$
$$= \prod_{i:c(q_i;d)>0} \frac{p_s(q_i|d)}{p_u(q_i|d)} \bullet \prod_i p_u(q_i|d) \quad (4)$$

A model $p_s(q_i|d)$ used for "seen" words that occur in the document, and a model $p_u(q_i|d)$ for "unseen" words that do not. The probability of a query $q$ can be written in terms of these models as above, where $c(w;d)$ denotes the count of word w in d.

We define modifying algorithm as follow:

Assume that video V's text information is T. L(V) is the V's value of relevance to a query q based on language model. PM(V) is the playnum of video V. P(V) is the proportion of video V's playnum in the sum of all videos'. $a$ is the effect factor of P(V), and b is the effect factor of VR(V). M(V) is the modified ultimate value of V. The modified ultimate value of a video V is given as follows:

$$L(V) = p(q|T) \quad (5)$$
$$P(V) = \frac{PM(V)}{\sum_i PM(V_i)} \quad (6)$$
$$M(V) = (1-a-b) \bullet L(V) + a \bullet P(V) + b \bullet VR(V)$$
$$a>0, b>0, a+b<1 \quad (7)$$

The video search engine need different effect factors a and b for different online video sites to gain the best performance due to different building algorithms adopted by online video sites. Thus, we need to train the effect factors for the video site.

The effect factors training takes three steps:

- Choose enough queries for training (at least 50 is the TREC's standard).
- Use different (a, b) to calculate every query's precision and calculate the average precision.
- Find the values of a and b when the average precision is maximum.

The precision is defined as follow:

$$\text{precision} = \frac{\text{the number of relevant videos for query}}{\text{the number of all videos in search result}} \quad (8)$$

## 6. EXPERIMENTS AND RESULTS

In this Section, we present the experimental results of the proposed system based on our video processing method, including the performance analysis of our system, and some comparisons with the traditional method which is adopted by the commercial video search engines, in terms of the precision.

### 6.1 Training Effect Factors

In order to make the different languages not affecting the experimental results, we used tudou [6], the Chinese largest online video site, as experimental subject.

Moreover, in order to decrease the effect of different query to experimental result. We randomly chose 60 queries for training effect factors of tudou [6]. The effect factor a and b were both selected from 0.1, 0.2, 0.3, 0.4, 05, 0.6, 0.7, 0.8. First of all, we fixed the values of a and b which were selected from 0.1, 0.2, 0.3, 0.4, 05, 0.6, 0.7, 0.8 and a+b<1. And then, we used every experimental query for search and we got 60 precisions from search results. Third, we calculated the average precision of the 60 precisions. Furthermore, we changed the values of a and b different from above and repeated the abovementioned steps until all of the values were chosen. The average precisions of different a and b are listed in Table 1.

**Table 1. The average precisions for different a and b**

| a | b | Average Precision | a | b | Average Precision |
|---|---|---|---|---|---|
| 0.1 | 0.1 | 0.8364 | 0.3 | 0.4 | 0.3086 |
| 0.1 | 0.2 | 0.8710 | 0.3 | 0.5 | 0.2083 |
| 0.1 | 0.3 | 0.9011 | 0.3 | 0.6 | 0.1701 |
| 0.1 | 0.4 | 0.9345 | 0.4 | 0.1 | 0.4017 |
| 0.1 | 0.5 | 0.7326 | 0.4 | 0.2 | 0.2545 |
| 0.1 | 0.6 | 0.5725 | 0.4 | 0.3 | 0.3104 |
| 0.1 | 0.7 | 0.3382 | 0.4 | 0.4 | 0.1452 |
| 0.1 | 0.8 | 0.2463 | 0.4 | 0.5 | 0.1597 |
| 0.2 | 0.1 | 0.7552 | 0.5 | 0.1 | 0.2146 |
| 0.2 | 0.2 | 0.6193 | 0.5 | 0.2 | 0.1562 |
| 0.2 | 0.3 | 0.5710 | 0.5 | 0.3 | 0.1573 |
| 0.2 | 0.4 | 0.5850 | 0.5 | 0.4 | 0.1070 |
| 0.2 | 0.5 | 0.3579 | 0.6 | 0.1 | 0.1955 |
| 0.2 | 0.6 | 0.2434 | 0.6 | 0.2 | 0.1008 |

| 0.2 | 0.7 | 0.2198 | 0.6 | 0.3 | 0.0621 |
| 0.3 | 0.1 | 0.5187 | 0.7 | 0.1 | 0.0368 |
| 0.3 | 0.2 | 0.4032 | 0.7 | 0.2 | 0.0376 |
| 0.3 | 0.3 | 0.4147 | 0.8 | 0.1 | 0.0210 |

From the Table 1, we can see the average precision is maxium when the a is 0.1 and b is 0.4. Thus, effect factors for tudou [6] are 0.1 and 0.4.

## 6.2 Evaluation of Method Based on Pavideoge

In this section, we evaluate the performance of the video processing method based on pavideoge. We compared our system's performance to tudou [6] with the system based on traditional method which is used by commercial video search engines.

In order to set up a consistent data collection for further evaluation and comparison and eliminate the impact of the different data set on comparing result, we first used about 80,000 videos and videos' webpages from tudou [6], which had randomly crawled in our previous research. Second, we built another video system based on traditional method. Both systems utilize lemur [15] to build index based on language model and use the same data set as mentioned in the first step.

In order to decrease the effect of particular query to comparing result, we randomly chose 6 groups of test queries and each group had 50 test queries. We used 50 queries of each group to test our system and the system based on traditional method and calculated the average precision of each group. The values of the average precisions of each group are listed in Table 2. And the comparing result is shown in Figure 10.

**Table 2. The average precisions of each group**

| Group | Use Pavideoge | No Use Pavideoge | Percentage of Improvement |
|---|---|---|---|
| 1 | 0.9271 | 0.8438 | 9.232% |
| 2 | 0.9106 | 0.8732 | 4.283% |
| 3 | 0.8491 | 0.8507 | -0.188% |
| 4 | 0.8788 | 0.8673 | 1.326% |
| 5 | 0.8480 | 0.7392 | 14.719% |
| 6 | 0.8415 | 0.8410 | ≈ 0% |

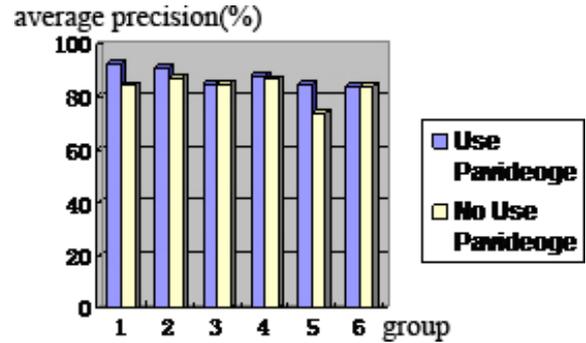

**Figure 10. The comparsion of the average precisions of each group by two different methods**

From the result, we can see that the average precisions in test group 1, 2, 4 and 5 are enhanced. Particularly, the average precision is increased by 14.719%. The average precision in test group 3 is decreased by 0.188% and in group 6, the average precision is almost not improved, but their effect is small compared with the other test groups. The average percentage of improvement is 4.895%, thus, we can see our new processing video method is better than traditional method in enhancing the search precision.

## 7. CONCLUSION AND FUTURE WORK

In this paper, we presented a novel video processing method in video search engine. The traditional method adopted by commercial video search engines cannot effectively handle the immanent attributes of video: requiring enormous storage space, being independent, expressing information hiddenly. The traditional means uses the video's title sometimes and tag instead of the video's content. Unfortunately, the title and tag are not enough to present the video. In our method, we first create the new data type—pavideoge for dealing with the video's inner properties, which includes: real link, videorank, text information and playnum. And then we use the pavideoge to modify the search result. The real link is used to handle the video's enormous storage space by simulating the actions of server of online video site and providing true hyperlink of the video in video database. The videorank is used to handle to independence of video by revealing the relevance, which is fixed when an online video site is built, existing between videos. Furthermore, we use text information to build language model index, and generate the final search result modified by videorank and playnum. We propose the modifying algorithm, which uses effect factors to balance the influences to search result from the users' perspective and the online video site's perspective. To evaluate the proposed video processing method, we conducted experiments on Chinese biggest online video site. The experimental results demonstrated promising performance of our method in terms of enhancing the precision of the search result, compared with traditional method adopted by commercial video search engine.

In the future, we will study whether the best effect factors of some similar online video sites are the constants. Furthermore, we will further reveal what are the determinant factors which make those effect factors constant and the true relationship between those similar online video sites.

## 8. REFERENCES

[1] Video. http://en.wikipedia.org/wiki/Video.

[2] Video Search Engine. http://en.wikipedia.org/wiki/Video_search_engine.

[3] http://video.google.cn/.

[4] http://video.baidu.com/.

[5] http://v.sogou.com/.

[6] http://www.tudou.com.

[7] S. Brin and L. Page, "The anatomy of a large-scale hypertextual Web search engine", Proceedings of the Seventh International World Wide Web Conference, 1998.

[8] L. Page, S. Brin, R. Motwani and T. Winograd, "The PageRank Citation Ranking: Bringing order to the Web", http://infolab.stanford.edu/~backrub/pageranksub, 1998.

[9] Christopher D. Manning, Prabhakar Raghavan, and Hinrich Schütze, An Introduction to Information Retrieval, Cambridge University Press, July 12, 2008.

[10] PageRank. http://en.wikipedia.org/wiki/PageRank.

[11] Chris Ridings, and Mike Shishigin, PageRank Uncovered, 2002.

[12] Taher H. Haveliwala, "Efficient Computation of PageRank", October 18, 1999.

[13] Chengxiang Zhai, John Lafferty,"A Study of Smoothing Methods for Language Models Applied to Ad Hoc Information Retrieval", SIGIR'01, New Orleans, Louisiana, USA, September 9-12, 2001.

[14] Chengxiang Zhai, John Lafferty," A Study of Smoothing Methods for Language Models Applied to Information Retrieval", ACM Transactions on Information Systems, Vol. 22, No. 2, April 2004, Pages 179–214.

[15] Lemur. http://www.lemurproject.org/